\documentclass[aps,prl,twocolumn,superscriptaddress]{revtex4-1}

\usepackage[hidelinks]{hyperref}

\usepackage{physics}
\usepackage{balance}
\usepackage{suffix}
\usepackage{mathtools}
\usepackage[utf8]{inputenc}
\usepackage{booktabs}
\usepackage{cases}
\usepackage[multiple]{footmisc}
\usepackage{dcolumn}
\usepackage{color,soul}
\usepackage{rotating}
\usepackage{perpage}
\usepackage{siunitx}
\usepackage{xcolor}
\definecolor{darkcyan}{RGB}{0,139,139}
\hypersetup{
	colorlinks,
	linkcolor={red},
	citecolor={blue},
	urlcolor={blue}
}
\usepackage{soul}
\usepackage{amsmath}
\usepackage{tikz}
\usepackage[T1]{fontenc}
\usepackage{etoolbox}
\usepackage{graphics}
\usepackage{siunitx}
\usepackage{float}	
\usepackage{collref}
\usepackage{multirow}
\usepackage{mathtools}
\usepackage{bm}
\usepackage{url}
\usepackage{pifont}
\usepackage{balance}

\usepackage{amssymb}
\usepackage{mathtools}
\usepackage{amsmath}
\usepackage{bm}
\usepackage{siunitx}
\usepackage{graphicx}
\usepackage{float}
\usepackage{multirow}
\usepackage{booktabs}
\usepackage{tikz}
\usepackage{tikz-3dplot}
\usepackage{pgfplots}
\pgfplotsset{compat=1.17}

\usepackage{tikz}
\usepackage{tikz-3dplot}
\usepackage{accents}

\makeatletter
\newcommand{\doublewidetilde}[1]{{%
		\mathpalette\double@widetilde{#1}%
}}
\newcommand{\double@widetilde}[2]{%
	\sbox\z@{$\m@th#1\widetilde{#2}$}%
	\ht\z@=.9\ht\z@
	\widetilde{\box\z@}%
}
\makeatother

\usepackage{etoolbox,lipsum}

\begin{document}

\title{Symmetry-selective nonrelativistic spin splitting in antiferromagnets\\ driven by coherent phonons}

\author{Sangeeta Rajpurohit}
\email{rajpurohit1@llnl.gov}
\affiliation{Material Science Division, Lawrence Livermore National Laboratory, CA 94550, USA}

\author{Mohsen Yarmohammadi}
\affiliation{Department of Physics, Georgetown University, Washington DC 20057, USA}

\author{Sheikh Rubaiat Ul Haque}
\affiliation{Department of Applied Physics, Stanford University, Stanford, CA 94305, USA}
\affiliation{Department of Materials Science and Engineering, Stanford University, Stanford, CA 94305, USA}
\affiliation{Stanford Institute for Materials and Energy Sciences, SLAC National Accelerator Laboratory, Menlo Park, CA 94025, USA}

\author{Tony F. Heinz}
\affiliation{Department of Applied Physics, Stanford University, Stanford, CA 94305, USA}
\affiliation{Stanford Institute for Materials and Energy Sciences, SLAC National Accelerator Laboratory, Menlo Park, CA 94025, USA}
\affiliation{Stanford PULSE Institute, SLAC National Accelerator Laboratory, Menlo Park, CA 94025, USA}

\author{Aaron M. Lindenberg} 
\affiliation{Department of Materials Science and Engineering, Stanford University, Stanford, CA 94305, USA}
\affiliation{Stanford Institute for Materials and Energy Sciences, SLAC National Accelerator Laboratory, Menlo Park, CA 94025, USA}
\affiliation{Stanford PULSE Institute, SLAC National Accelerator Laboratory, Menlo Park, CA 94025, USA}

\author{Tadashi Ogitsu}
\affiliation{Material Science Division, Lawrence Livermore National Laboratory, CA 94550, USA}

\date{\today}

\begin{abstract}
Nonrelativistic spin splitting (NRSS) in antiferromagnets (AFMs) enables magnetization-free spin polarization for ultrafast spintronics.
Here, we demonstrate that coherent phonons can dynamically induce and control NRSS in collinear AFMs. Excitation of $\Gamma$-point
infrared-active phonons lifts the spin degeneracy of the ground state, while the residual sublattice-connecting symmetries determine
the momentum-space form of the induced splitting. Because the relevant infrared modes couple to orthogonal in-plane light polarizations,
distinct spin-split phases can be selectively activated by the polarization of the driving field. Using first-principles calculations
for MnPS$_3$, we show that a mode that breaks all sublattice-connecting symmetries induces an $s$-wave spin-split state with
$\Delta(\Gamma)\neq0$, whereas a symmetry-distinct mode that preserves a sublattice-connecting mirror symmetry generates a $d$-wave
altermagnetic state with $\Delta(\Gamma)=0$. In both cases, the spin splitting grows linearly with the phonon amplitude and reverses
the induced spin polarization when the displacement is reversed. Our results establish coherent lattice driving as a direct,
polarization- and mode-selective route to dynamically induce distinct NRSS phases in AFMs.
\end{abstract}
\maketitle

{\allowdisplaybreaks
\textit{Introduction}---The recent discovery of altermagnetism~\cite{Smejkal2022_1,Smejkal2022_2,Mazin2022,Hayami2019,Smejkal2022_3,Krempasky2024,Reimers2024}
has renewed our interest in unconventional nonrelativistic spin-splitting in
antiferromagnets (AFMs)~\cite{Zelezny2014, Wadley2016, Bodnar2018, Olejnik2018, Salemi2019}.
Unlike the exchange spin-splitting in ferromagnets \cite{Neel1953,Chernyshov2009, Miron2010, Fang2011, Miron2011, Kurebayashi2014, Ciccarelli2016},
which is accompanied by a net magnetization, or the Rashba and Dresselhaus
spin-splittings arising from spin-orbit coupling (SOC) \cite{Rashba1960,Dresselhaus1955,Manchon2015,Manchon2019,Trier2022},
nonrelativistic spin-splitting (NRSS) occurs in compensated magnetic systems without
requiring either net magnetization or relativistic effects \cite{Yuan2021,Yuan2020}. From
a symmetry perspective, AFMs exhibiting NRSS are characterized by the simultaneous absence
of the translation-spin-reversal ($Ut$) and inversion-time-reversal ($P\mathcal{T}$) symmetries,
where $U$ reverses the spin, $t$ denotes translation, $\mathcal{T}$ is time reversal, and $P$ is
spatial inversion. Equivalently, the two opposite-spin sublattices are related by neither translation nor inversion.
Any compensated AFM satisfying these symmetry conditions can, in principle, exhibit NRSS \cite{Yuan2021,Yuan2020}.

\begin{figure}[t]
\centering
\includegraphics[width=1\linewidth]{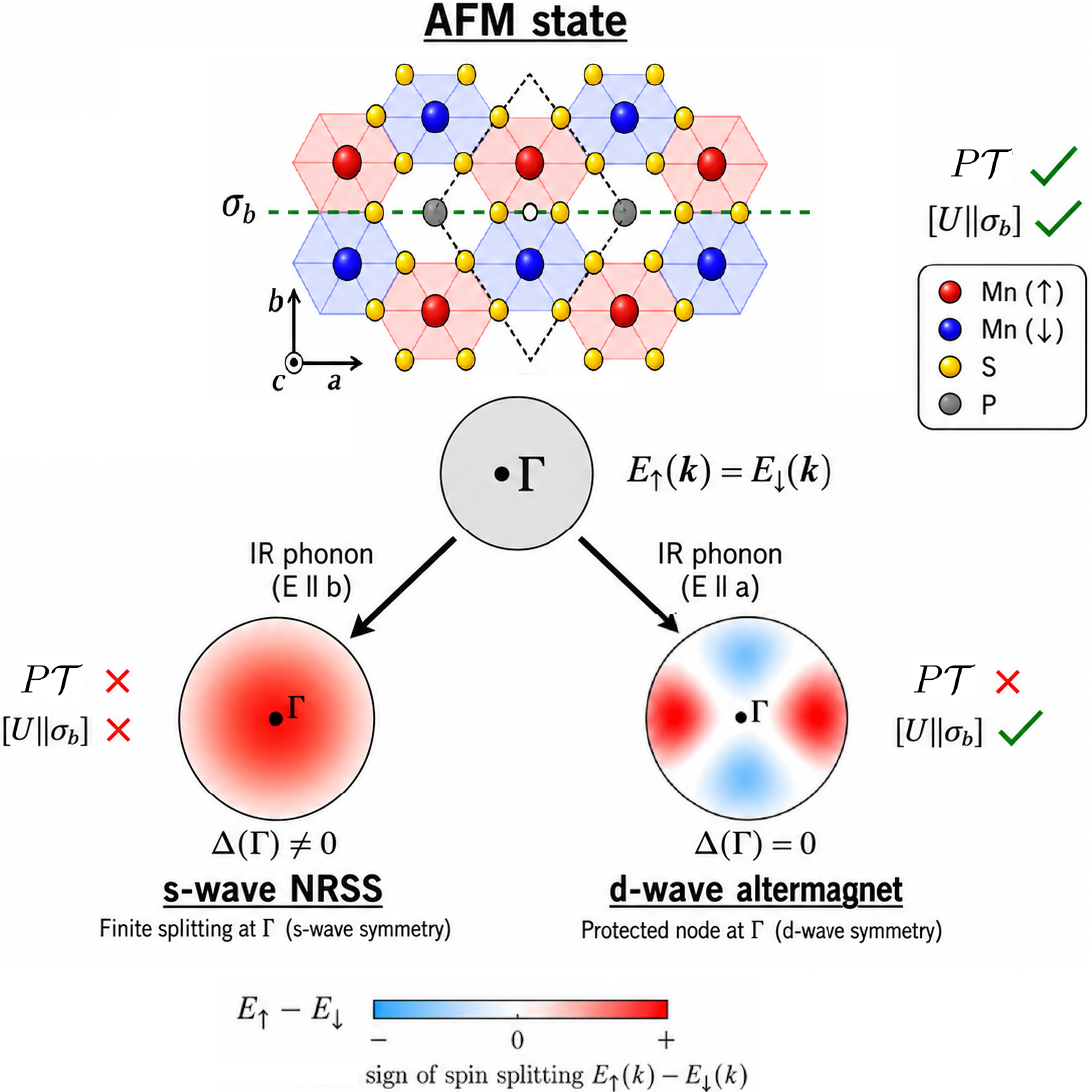}
\caption{\textbf{Phonon-driven, symmetry-selective nonrelativistic spin splitting (NRSS) in MnPS$_3$.}
In the AFM ground state (top), the opposite-spin Mn sublattices are related by both
$P\mathcal{T}$ and the sublattice-connecting symmetry $[U\|\sigma_b]$,
enforcing spin degeneracy, $E_\uparrow(\mathbf{k})=E_\downarrow(\mathbf{k})$.
Excitation of the $A_u$ phonon ($E\parallel b$) breaks both
$P\mathcal{T}$ and $[U\|\sigma_b]$, producing an $s$-wave NRSS with
$\Delta(\Gamma)\neq0$ (bottom-left). In contrast, excitation of the $B_u$ phonon
($E\parallel a$) breaks $P\mathcal{T}$  but preserves $[U\|\sigma_b]$,
resulting in a symmetry-protected $d$-wave spin splitting with
$\Delta(\Gamma)=0$, characteristic of an altermagnetic phase (bottom-right).
Checkmarks and crosses denote preserved and broken symmetries,
respectively.}
\label{fig:fig_1}
\end{figure}

Collinear altermagnets constitute a symmetry-protected subset of this broader
class~\cite{Gonzalez2021,Bai2022,Karube2022,Lee2024,Fedchenko2024,Krempasky2024, Osumi2024}.
In addition to the absence of $Ut$ and $P\mathcal{T}$, the opposite-spin sublattices remain
connected by a proper or improper rotation $R$ \cite{Liu2022,Smejkal2022_1,Smejkal2022_2}. The
resulting $UR$ symmetry protects the spin degeneracy at the Brillouin-zone center, giving rise
to the characteristic momentum-dependent spin-splitting that changes sign across the Brillouin
zone while vanishing at $\Gamma$. By contrast, in the remaining compensated AFMs, no rotational
symmetry relates the opposite-spin sublattices. Consequently, the $UR$ symmetry is absent, the
zone-center degeneracy is no longer symmetry protected, and the spin-splitting is generally finite
at the $\Gamma$-point \cite{Yuan2024}.

A crucial question is whether NRSS can be induced and dynamically tuned in AFMs that do not intrinsically exhibit
NRSS in their equilibrium state. Most AFMs hosting NRSS, including altermagnetism, are stable structural phases
in which the spin-splitting is fixed by the underlying crystal symmetry \cite{Guo2023}. Modifying the NRSS in these
materials typically relies on static or quasi-static approaches such as chemical substitution \cite{Bandyopadhyay2025},
strain \cite{Bandyopadhyay2025,Zhang2025} and stacking engineering \cite{Wrzos2025}, or optical fields \cite{Rajpurohit2024,Yarmohammadi2025,Yarmohammadi2026}. In contrast, the dynamical
control of NRSS in AFMs remains largely unexplored. Terahertz~(THz)-field-driven lattice engineering has already proven successful
for controlling ferroelectricity~\cite{Xian2019}, superconductivity~\cite{Cavalleri2018}, topology \cite{Sie2019} and magnetism ~\cite{Disa2023,Haque2025,Yarmohammadi2025_2}
in quantum materials. Extending this approach to AFMs that can host NRSS would enable ultrafast switching, mode-selective
manipulation of the spin-splitting \cite{Wang2026}, and reversible control of spin-polarized electronic states. Infrared~(IR)-active phonons
provide a natural route by transiently modifying the crystal symmetry through coherent lattice distortions.  

In this Letter, we propose a symmetry-based design principle for phonon-induced NRSS in collinear AFMs~[Fig~\ref{fig:fig_1}]. 
Whenever a coherent $\Gamma$-point IR-active phonon removes the symmetry protecting the equilibrium spin
degeneracy without restoring an equivalent sublattice-connecting symmetry, NRSS emerges, with its
momentum-space symmetry dictated by the residual crystal symmetries. We demonstrate this concept in
the layered van der Waals AFM MnPS$_3$, whose equilibrium magnetic point group $2/m'$ preserves
$P\mathcal{T}$ symmetry and enforces spin degeneracy throughout the Brillouin zone. 

Using first-principles calculations, we show that two nearly degenerate THz-frequency IR-active phonons ($A_u$ and $B_u$) induce qualitatively
different forms of NRSS. One phonon removes all symmetries connecting the opposite-spin sublattices,
generating an $s$-wave spin splitting with $\Delta E_{\uparrow\downarrow}(\Gamma)\neq0$. The other preserves a sublattice-connecting
mirror symmetry, producing a symmetry-protected $d$-wave spin splitting with $\Delta E_{\uparrow\downarrow}(\Gamma){=}0$, characteristic
of an altermagnetic phase. The spin splitting reverses sign when the phonon displacement is reversed.
These results establish coherent phonon excitation as a symmetry-selective route
for inducing and controlling distinct forms of NRSS in AFMs, with altermagnetism emerging as one symmetry-protected realization.
Crucially, both phonon modes are IR active and therefore couple directly to light, while their orthogonal in-plane polarizations
allow the symmetry of the induced spin splitting to be selected by the polarization of the driving field.

\textit{Calculation setup}---We perform DFT+$U$ calculations for bulk MnPS$_3$ using Quantum ESPRESSO \cite{Giannozzi2009,Giannozzi2017}
with the Perdew-Burke-Ernzerhof (PBE) exchange-correlation functional \cite{Perdew1996} and 
pseudopotentials \cite{Bloechl1994}. A plane-wave cutoff energy of 80 Ry and a $4\times4\times2$ Monkhorst-Pack
$k$-point grid are employed. The on-site Coulomb interaction for the Mn $d$ orbitals is treated using $U=2.5$ eV. We
use a primitive unit cell containing 2 Mn, 2 P, and 6 S atoms, with lattice parameters $a=b=6.07$ \AA, and $c=6.76$ \AA \cite{Haque2025}.
The zone-center phonon frequencies and eigenvectors are taken from our previous first-principles
phonon calculations~\cite{Haque2025}. We use these modes to construct frozen-phonon distortions, computing
the electronic and spin structure self-consistently as a function of the displacement amplitude. Electronic
self-consistency is achieved with a total-energy convergence threshold of $10^{-7}$ Ry. All calculations are performed
for the AFM spin configuration of the Mn atoms. The dependence of the results on the Hubbard interaction parameter
$U$ has been systematically examined and is presented in the Supplemental Material~\cite{SM}, demonstrating the
robustness of our conclusions. Because NRSS does not require relativistic effects, SOC is neglected throughout.

\begin{figure}[t]
\centering
\includegraphics[width=1\linewidth]{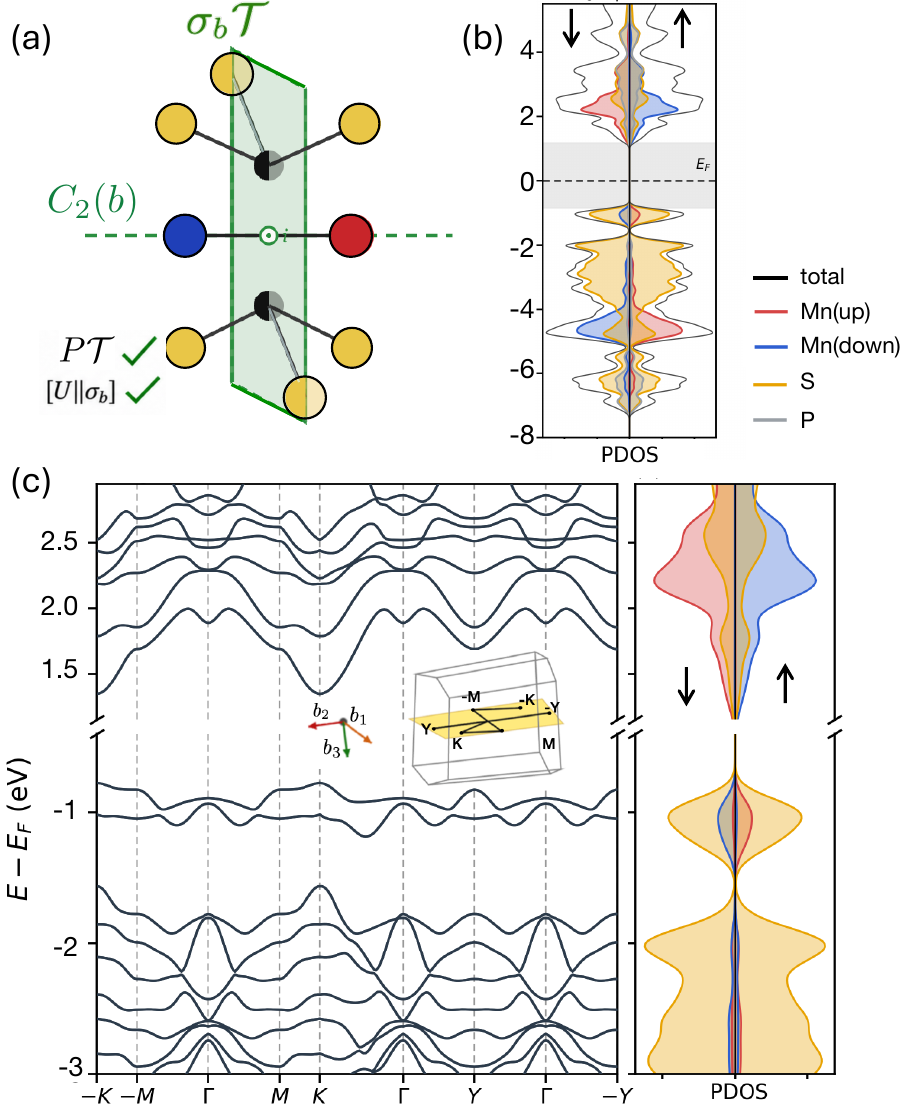}
\caption{\textbf{Ground-state symmetry and electronic structure of MnPS$_3$.}
(a) Two equilibrium magnetic symmetry operations governing nonrelativistic spin
splitting: the antiunitary mirror $\sigma_b\mathcal{T}$ (green $ac$-plane) and the
two-fold rotation $C_2(b)$ (green dashed axis). Blue and red spheres are Mn atoms with
opposite spins, yellow and black represent S and P.  Both $P\mathcal{T}$ and $[U\|\sigma_b]$ enforce spin degeneracy
at every $\mathbf{k}$. (b) Orbital- and spin-projected DOS: total (black), Mn(up) (red), Mn(down) (blue),
S (yellow), P (grey). (c) Spin-resolved band structure along
$\text{-}K-\text{-}M-\Gamma-M-K-\Gamma-Y-\Gamma-\text{-}Y$ (inset: Brillouin zone and path),
with the projected DOS alongside. Bands are spin-degenerate everywhere.}
\label{fig:fig_2}
\end{figure}

\begin{figure*}[t]
\centering
\includegraphics[width=0.9\linewidth]{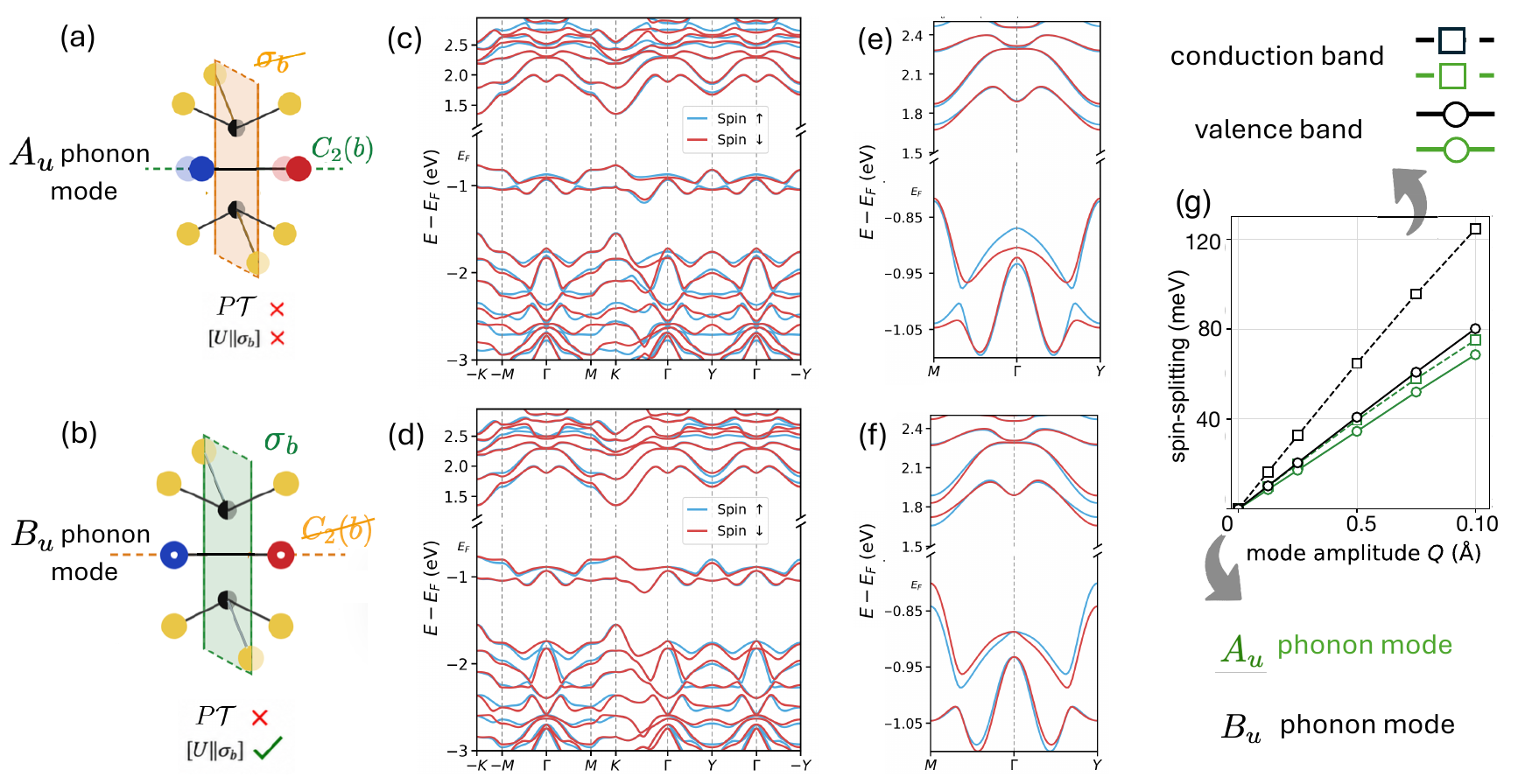}
\caption{\textbf{Symmetry analysis and band structure of MnPS$_3$ under IR-active phonon distortions.}
(a-b) Side view of the two IR-active phonon modes.
The shaded plane and dashed line denote the $\sigma_b$ mirror plane and $C_2(b)$ rotation axis, respectively.
Blue/red spheres represent Mn atoms with opposite spins, while yellow and black spheres denote S and P atoms.
Faded and solid Mn spheres indicate equilibrium and displaced positions. The $A_u$ mode (a), involving Mn
displacement along $b$, preserves $C_2(b)$ but breaks $\sigma_b\mathcal{T}$, whereas the $B_u$ mode (b),
involving Mn displacement along $c$, preserves $\sigma_b\mathcal{T}$ but breaks $C_2(b)$. (c and d)
Spin-resolved band structures of MnPS$_3$ in the $A_u$-distorted (c) and $B_u$-distorted (d) states.
(e-f) Zoomed-in view of the top of the valence and bottom of the conduction band for $A_u$-mode with
nonzero spin-splitting through $\Gamma$ (e) and $B_u$-mode with finite spin-splitting away from $\Gamma$
that vanishes at $\Gamma$ (f).  (g) Linear scaling of the spin-splitting with phonon amplitude for
valence (circles) and conduction (squares) bands. Green and black correspond to $A_u$ and $B_u$ modes, respectively.
The splitting reverses sign upon reversing the displacement.}
\label{fig:fig_3}
\end{figure*}

\textit{Symmetry analysis}---The AFM ground state of MnPS$_3$ below $T_N$ has a crystal point group
$C_{2h}$. MnPS$_3$ orders antiferromagnetically within the $ab$-plane [Fig.~\ref{fig:fig_1}, top] and has
ferromagnetic coupling between layers along the $c$ direction, forming a magnetic point group $2/m'$. While the
crystal point group $C_{2h}$ preserves the inversion symmetry, the magnetic
point group $2/m'$ breaks the inversion symmetry $P$, as shown in Fig.~\ref{fig:fig_2}(a). The magnetic group
$2/m'$ contains four symmetry operations that preserve the magnetic ground
state: the identity $E$, the unitary two-fold rotation $C_2(b)$ about the $b$
axis, and the antiunitary operations $P\mathcal{T}$ and $\sigma_b\mathcal{T}$,
where $\sigma_b$ denotes the mirror plane perpendicular to the $b$ axis (i.e.
the $ac$-plane) and $\mathcal{T}$ is time reversal. The combination
$P\mathcal{T}$ enforces Kramers' degeneracy at every $k$-point in the Brillouin
zone, as confirmed by the spin-projected band structure in Fig.~\ref{fig:fig_2}(c).
The top of the valence band in the energy window $-1.5$ to $0.5$ eV consists of antibonding states
largely involving $p$-orbitals of the S atoms with some contribution from Mn $d$-states
[Figs.~\ref{fig:fig_2}(b) and \ref{fig:fig_2}(c)]. The corresponding bonding states are formed by
the lower valence band. Clearly, the spin degeneracy is intact at every $k$-point, reflecting the preserved
$P\mathcal{T}$ symmetry in the AFM ground state.

Here we show that selective excitation of symmetry-distinct phonon modes in collinear AFMs can engineer
both the emergence and the momentum-space symmetry of NRSS~[see Fig.~\ref{fig:fig_1}]. This mechanism requires
a spin-degenerate AFM whose equilibrium spin degeneracy is protected by a symmetry that can be broken by a zone-center
infrared-active phonon. Because $\Gamma$-point phonons preserve lattice translations, they cannot lift spin degeneracy
protected by translation-related symmetries. In contrast, a symmetry-allowed IR-active $\Gamma$-point phonon can break
antiunitary symmetries containing inversion, such as the $P\mathcal{T}$ symmetry of MnPS$_3$, thereby lifting
the spin degeneracy. Furthermore, the phonon distortion must remove the sublattice-connecting symmetry without
leaving any equivalent sublattice-connecting operation intact. The momentum-space structure of the resulting
spin splitting is then determined by the residual sublattice-connecting symmetries. These symmetry requirements
establish a concrete design principle for identifying AFMs in which coherent phonon excitation can induce NRSS
and engineer its momentum-space symmetry.

Several low-frequency IR-active phonons have been reported in AFM MnPS$_3$~\cite{Vaclavkova2020,Rao2024,Haque2025}.
We classify the three modes near 4.5~THz according to the irreducible representations of the parent paramagnetic
crystal structure by computing the characters $\chi_g=\langle\mathbf{e}|g|\mathbf{e}\rangle$ of the phonon
eigenvectors $\mathbf{e}$ under each $C_{2h}$ symmetry operation $g$: the modes at 4.36 and 4.49~THz transform
as $B_u$, while the mode at 4.47~THz transforms as $A_u$. Although inversion symmetry is formally broken
below $T_\mathrm{N}$, the resulting symmetry lowering has a negligible effect on the phonon frequencies and
eigenvectors. We therefore retain the parent-structure irreducible-representation labels for clarity. All three
modes are odd under inversion in the parent structure and are therefore IR-active. We focus on the two modes
whose polar axes lie in the $ab$ plane, as these couple to normally incident THz radiation in the layered geometry;
the remaining mode is polarized along $c$ and is not considered further. Freezing in either mode reduces the crystal
symmetry in a distinct way, leading to different forms of NRSS. As shown below, the $A_u$ mode removes all
sublattice-connecting symmetries, giving rise to an $s$-wave spin splitting, whereas the $B_u$ mode preserves
a sublattice-connecting mirror symmetry, producing the symmetry-protected $d$-wave spin splitting characteristic of
an altermagnetic phase.

The 4.47~THz $A_u$ phonon mode is characterized by Mn displacements along $\pm\hat{b}$, opposite to those of the
S and P atoms, resulting in a polar distortion along the $C_2$ axis, as shown in Fig.~\ref{fig:fig_3}(a). This
distortion breaks the antiunitary symmetries $\sigma_b\mathcal{T}$ and $P\mathcal{T}$, leaving only the unitary
$C_2(b)$ symmetry intact. Because $C_2(b)$ maps each spin sublattice onto itself, no symmetry operation remains to
connect the two opposite-spin Mn sublattices, and a finite spin-splitting, $\Delta E_{\uparrow\downarrow}(\Gamma)\neq0$,
emerges at the $\Gamma$ point. Figures~\ref{fig:fig_3}(c) and \ref{fig:fig_3}(e) show the electronic band structure for
the frozen $A_u$ mode, revealing a pronounced spin-splitting at $\Gamma$. For a phonon amplitude of $Q = 0.05$~\AA,
the topmost valence and lowest conduction bands exhibit spin splittings of approximately $50$~meV, while the
maximum spin splitting among the near-Fermi bands, within a few bands of the valence and conduction edges, reaches
$\sim100$~meV.

The $A_u$-distorted state  belongs to the broader class of nonrelativistic spin-split AFMs recently identified
by Yuan et al. \cite{Yuan2024}, in which no proper or improper rotation relates the opposite-spin sublattices and
the zone-center degeneracy is lifted. The $A_u$ phonon therefore provides a dynamical route to this
$\Gamma$-split NRSS phase, complementing the symmetry-protected $d$-wave altermagnetic spin-splitting
induced by the $B_u$ mode, as discussed next.

\begin{figure}[t]
\includegraphics[width=1\linewidth]{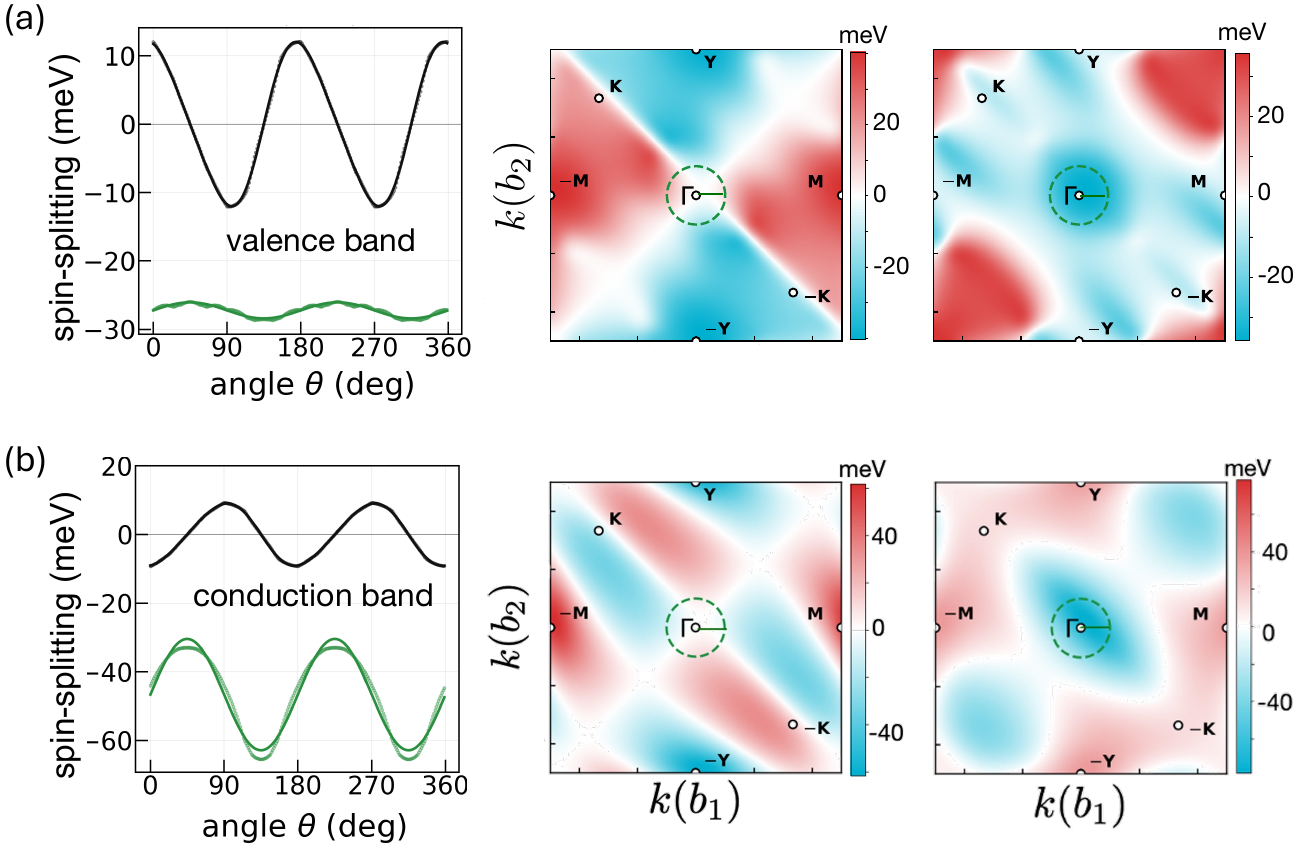}
\caption{\textbf{Symmetry of the phonon-induced spin-splitting.}
Angular profile of the spin-splitting $E_\uparrow(\mathbf{k}){-}E_\downarrow(\mathbf{k})$ for the $A_u$ (green)
and $B_u$ (black) modes of amplitude $A=0.05$~\AA{} on a circle of radius $|\mathbf{k}|= 0.1$~$(2\pi/a)$ around
$\Gamma$ for the top of the valence band (a) and the bottom of the conduction band (b). The 2D
Brillouin-zone spin-splitting for the corresponding $A_u$-distorted (middle) and $B_u$-distorted
(right) states. Red and blue indicate positive and negative sign. The $A_u$ mode gives an isotropic
$s$-wave splitting around $\Gamma$; the $B_u$ mode gives a $d$-wave splitting with two nodal lines through $\Gamma$.
Green and black indicate $A_u$ and $B_u$ modes. Open and closed symbols indicate the two-component Fourier fit.
The $A_u$ mode has a dominant $s$-wave pattern and the $B_u$ mode has a pure $\cos 2\theta$ $d$-wave with four 
zero crossings per revolution.}
\label{fig:fig_4}
\end{figure}

The 4.49~THz $B_u$ mode consists of parallel displacements of both Mn atoms
along $\hat{c}$ within the $\sigma_b$ mirror plane [Fig.~\ref{fig:fig_3}(b)].
This distortion breaks $P\mathcal{T}$ and $C_2(b)$ but preserves the mirror
symmetry $\sigma_b$, an improper rotation that connects the two opposite-spin
Mn sublattices. Consequently, the surviving spin-group operation
$[U\|\sigma_b]$ (equivalently, the magnetic operation $\sigma_b\mathcal{T}$)
remains the sublattice-connecting $[U\| R]$ symmetry with $R=\sigma_b$, which
protect the spin degeneracy at $\Gamma$. As a result, the induced spin
splitting retains the characteristic $d$-wave symmetry of an altermagnetic
phase, vanishing at all high-symmetry points. Figures~\ref{fig:fig_3}(d) and \ref{fig:fig_3}(f)
show that the spin degeneracy is lifted away from $\Gamma$, while
$\Delta E_{\uparrow\downarrow}(\Gamma)=0$ remains protected by the retained $[U\|\sigma_b]$ symmetry.
Moreover, the spin splitting changes sign between the $\Gamma$-M and
$\Gamma$-Y directions, consistent with $d$-wave symmetry.

Including SOC in the frozen-phonon calculations does not qualitatively alter the results. Fully relativistic
calculations reproduce both the $s$-wave ($A_u$) and the $d$-wave ($B_u$) spin splittings to within $\sim10\%$.
The $A_u$-induced splitting at $\Gamma$ remains nearly unchanged, while the $B_u$ node at $\Gamma$ is lifted by
only a few meV, far smaller than the $d$-wave antinode. These results confirm that the mode-selective NRSS is
predominantly nonrelativistic in origin (see Supplemental Material).

The spin-splitting increases linearly for both the valence- and conduction-band edges with the phonon-mode
amplitude, as shown in Fig.~\ref{fig:fig_3}(g). To further analyze the symmetry of the induced spin-splitting, we
calculate the two-dimensional momentum-resolved spin-splitting of the valence-band maximum and conduction-band
minimum, as shown in Fig.~\ref{fig:fig_4}. Figures~\ref{fig:fig_4}(a) and \ref{fig:fig_4}(b) display the angular
dependence of the spin-splitting, $E_{\uparrow}(\textbf{k})-E_{\downarrow}(\textbf{k})$, around the $\Gamma$-point for
both phonon modes. The $A_u$ mode produces a predominantly $s$-wave spin-splitting, with the same sign in all momentum
directions. In contrast, the $B_u$ mode generates a characteristic $d$-wave spin-splitting with two nodal lines passing
through the $\Gamma$ point, along the $K$-$\Gamma$-$K$ direction and the orthogonal direction, across which the
spin-splitting changes sign. Recent time-resolved SHG-RA measurements on THz-driven MnPS$_3$~\cite{Haque2025}
are in line with our predictions: since SHG is magnetically induced and sensitive to the spin-resolved electronic
structure~\cite{Fiebig2005}, the strong polarization-sensitive response of the SHG-RA signals at the
$\sim 4.5$~THz $A_u$ and $B_u$ modes reflects the polarization-controlled symmetry breaking proposed here. 

We demonstrate that coherent excitation of infrared-active phonons can generate nonrelativistic
anisotropic spin splitting in conventional AFMs. More generally, our results establish a symmetry-based
design principle for phonon-induced NRSS in spin-degenerate collinear AFMs: the equilibrium spin degeneracy
must be protected by a symmetry that can be broken by a zone-center IR-active phonon, while the residual
sublattice-connecting symmetries determine the momentum-space symmetry of the induced spin splitting.
We illustrate this principle using MnPS$_3$, whose equilibrium $P\mathcal{T}$ symmetry enforces spin
degeneracy. Excitation of the $B_u$ phonon preserves the sublattice-connecting mirror symmetry $\sigma_b$,
producing the sign-changing momentum-space texture characteristic of altermagnetic ($d$-wave) spin splitting.
In contrast, the $A_u$ phonon removes all sublattice-connecting symmetries, lifting this constraint and
giving rise to an $s$-wave spin splitting with finite $\Delta E_{\uparrow\downarrow}(\Gamma)$. Although nearly degenerate in
frequency, the $A_u$ and $B_u$ phonons belong to different irreducible representations of the parent $C_{2h}$
point group and couple to orthogonal THz polarizations. Phonon displacements comparable to the $0.05$~\AA{} amplitude
considered here are experimentally accessible under resonant THz excitation \cite{Haque2025}, supporting the feasibility
of the predicted spin splitting. Since the induced spin splitting reverses sign
with the phonon displacement, a resonant THz drive that coherently oscillates these phonon coordinate
is expected to produce a spin splitting that oscillates at the phonon frequency. Consequently, polarization-selective
excitation is expected to independently drive the $s$- and $d$-wave NRSS phases. Furthermore, because the spin-splitting
amplitude scales linearly with the phonon amplitude, it can be continuously tuned by varying the intensity
of the driving field.

\textit{Conclusion}---In summary, this work establishes coherent infrared phonons as a robust, symmetry-selective
route to dynamically engineer nonrelativistic spin splitting in conventional
antiferromagnets. Beyond demonstrating optical control of the spin-splitting amplitude, we show that the driven
phonon mode determines the symmetry of the resulting spin texture, enabling reversible, ultrafast switching
between distinct nonrelativistic spin-split phases. These findings provide a general framework for
light-driven spin-symmetry engineering opens direct avenues for next-generation, low-power antiferromagnetic spintronics.

\textit{Acknowledgments}---S.\,R, S.\,R.\,U.\,H., T.\,F.\,H., A.\,M.\,L., and T.\,O. are supported by
the Computational Materials Sciences Program funded by the US Department of Energy,
Office of Science, Basic Energy Sciences, Materials Sciences and Engineering Division. 
M.\,Y. was supported by the Department of Energy, Office of Basic Energy Sciences,
Division of Materials Sciences and Engineering under Contract No. DE-FG02-08ER46542 for formal developments, 
results analysis, and manuscript writing. }

\let\oldaddcontentsline\addcontentsline
\renewcommand{\addcontentsline}[3]{}
\bibliography{ref.bib}

@article{Manchon2019,
	title = {{Current-induced spin-orbit torques in ferromagnetic and antiferromagnetic systems}},
	author = {Manchon, A. and \ifmmode \check{Z}\else \v{Z}\fi{}elezn\'y, J. and Miron, I. M. and Jungwirth, T. and Sinova, J. and Thiaville, A. and Garello, K. and Gambardella, P.},
	journal = {Rev. Mod. Phys.},
	volume = {91},
	issue = {3},
	pages = {035004},
	numpages = {80},
	year = {2019},
	month = {Sep},
	publisher = {American Physical Society},
	doi = {10.1103/RevModPhys.91.035004},
	url = {https://link.aps.org/doi/10.1103/RevModPhys.91.035004}
}

@article{Olejnik2018,
	author = {Kamil Olejník  and Tom Seifert  and Zdeněk Kašpar  and Vít Novák  and Peter Wadley  and Richard P. Campion  and Manuel Baumgartner  and Pietro Gambardella  and Petr Němec  and Joerg Wunderlich  and Jairo Sinova  and Petr Kužel  and Melanie Müller  and Tobias Kampfrath  and Tomas Jungwirth },
	title = {{Terahertz electrical writing speed in an antiferromagnetic memory}},
	journal = {Science Advances},
	volume = {4},
	number = {3},
	pages = {eaar3566},
	year = {2018},
	doi = {10.1126/sciadv.aar3566},
	URL = {https://www.science.org/doi/abs/10.1126/sciadv.aar3566}}

@article{Wadley2016,
	author = {P. Wadley  and B. Howells  and J. Železný  and C. Andrews  and V. Hills  and R. P. Campion  and V. Novák  and K. Olejník  and F. Maccherozzi  and S. S. Dhesi  and S. Y. Martin  and T. Wagner  and J. Wunderlich  and F. Freimuth  and Y. Mokrousov  and J. Kuneš  and J. S. Chauhan  and M. J. Grzybowski  and A. W. Rushforth  and K. W. Edmonds  and B. L. Gallagher  and T. Jungwirth },
	title = {{Electrical switching of an antiferromagnet}},
	journal = {Science},
	volume = {351},
	number = {6273},
	pages = {587-590},
	year = {2016},
	doi = {10.1126/science.aab1031},
	URL = {https://www.science.org/doi/abs/10.1126/science.aab1031}}

@article{Bodnar2018,
	author = {Bodnar, S. Yu. and {\v S}mejkal, L. and Turek, I. and Jungwirth, T. and Gomonay, O. and Sinova, J. and Sapozhnik, A. A. and Elmers, H. -J. and Kl{\"a}ui, M. and Jourdan, M.},
	date = {2018/01/24},
	doi = {10.1038/s41467-017-02780-x},
	id = {Bodnar2018},
	isbn = {2041-1723},
	journal = {Nature Communications},
	number = {1},
	pages = {348},
	title = {{Writing and reading antiferromagnetic Mn2Au by N{\'e}el spin-orbit torques and large anisotropic magnetoresistance}},
	url = {https://doi.org/10.1038/s41467-017-02780-x},
	volume = {9},
	year = {2018}}

@article{Trier2022,
	author = {Trier, Felix and No{\"e}l, Paul and Kim, Joo-Von and Attan{\'e}, Jean-Philippe and Vila, Laurent and Bibes, Manuel},
	date = {2022/04/01},
	doi = {10.1038/s41578-021-00395-9},
	id = {Trier2022},
	isbn = {2058-8437},
	journal = {Nature Reviews Materials},
	number = {4},
	pages = {258--274},
	title = {{Oxide spin-orbitronics: spin--charge interconversion and topological spin textures}},
	url = {https://doi.org/10.1038/s41578-021-00395-9},
	volume = {7},
	year = {2022}}

@article{Mazin2022,
	title = {{Editorial: Altermagnetism---A New Punch Line of Fundamental Magnetism}},
	author = {Mazin, Igor},
	collaboration = {The PRX Editors},
	journal = {Phys. Rev. X},
	volume = {12},
	issue = {4},
	pages = {040002},
	numpages = {3},
	year = {2022},
	month = {Dec},
	publisher = {American Physical Society},
	doi = {10.1103/PhysRevX.12.040002},
	url = {https://link.aps.org/doi/10.1103/PhysRevX.12.040002}
}

@article{Fedchenko2024,
	author = {Olena Fedchenko  and Jan Minár  and Akashdeep Akashdeep  and Sunil Wilfred D’Souza  and Dmitry Vasilyev  and Olena Tkach  and Lukas Odenbreit  and Quynh Nguyen  and Dmytro Kutnyakhov  and Nils Wind  and Lukas Wenthaus  and Markus Scholz  and Kai Rossnagel  and Moritz Hoesch  and Martin Aeschlimann  and Benjamin Stadtmüller  and Mathias Kläui  and Gerd Schönhense  and Tomas Jungwirth  and Anna Birk Hellenes  and Gerhard Jakob  and Libor Šmejkal  and Jairo Sinova  and Hans-Joachim Elmers },
	title = {{Observation of time-reversal symmetry breaking in the band structure of altermagnetic RuO$_2$}},
	journal = {Science Advances},
	volume = {10},
	number = {5},
	pages = {eadj4883},
	year = {2024},
	doi = {10.1126/sciadv.adj4883},
	URL = {https://www.science.org/doi/abs/10.1126/sciadv.adj4883}}

@article{Krempasky2024,
	author={Krempask{\'y}, J.
	and {\v{S}}mejkal, L.
	and D'Souza, S. W.
	and Hajlaoui, M.
	and Springholz, G.
	and Uhl{\'i}{\v{r}}ov{\'a}, K.
	and Alarab, F.
	and Constantinou, P. C.
	and Strocov, V.
	and Usanov, D.
	and others},
	title={{Altermagnetic lifting of {K}ramers spin degeneracy}},
	journal={Nature},
	year={2024},
	month={Feb},
	day={01},
	volume={626},
	number={7999},
	pages={517-522},
	issn={1476-4687},
	doi={10.1038/s41586-023-06907-7},
	url={https://doi.org/10.1038/s41586-023-06907-7}
}

@article{Osumi2024,
	title = {{Observation of a giant band splitting in altermagnetic {M}n{T}e}},
	author = {Osumi, T. and Souma, S. and Aoyama, T. and Yamauchi, K. and Honma, A. and Nakayama, K. and Takahashi, T. and Ohgushi, K. and Sato, T.},
	journal = {Phys. Rev. B},
	volume = {109},
	issue = {11},
	pages = {115102},
	numpages = {8},
	year = {2024},
	month = {Mar},
	publisher = {American Physical Society},
	doi = {10.1103/PhysRevB.109.115102},
	url = {https://link.aps.org/doi/10.1103/PhysRevB.109.115102}
}

@article{Bai2022,
	title = {{Observation of Spin Splitting Torque in a Collinear Antiferromagnet {R}u{O}$_2$}},
	author = {Bai, H. and Han, L. and Feng, X. Y. and Zhou, Y. J. and Su, R. X. and Wang, Q. and Liao, L. Y. and Zhu, W. X. and Chen, X. Z. and others},
	journal = {Phys. Rev. Lett.},
	volume = {128},
	issue = {19},
	pages = {197202},
	numpages = {6},
	year = {2022},
	month = {May},
	publisher = {American Physical Society},
	doi = {10.1103/PhysRevLett.128.197202},
	url = {https://link.aps.org/doi/10.1103/PhysRevLett.128.197202}
}

@article{Karube2022,
	title = {{Observation of Spin-Splitter Torque in Collinear Antiferromagnetic {R}u{O}$_{2}$}},
	author = {Karube, Shutaro and Tanaka, Takahiro and Sugawara, Daichi and Kadoguchi, Naohiro and Kohda, Makoto and Nitta, Junsaku},
	journal = {Phys. Rev. Lett.},
	volume = {129},
	issue = {13},
	pages = {137201},
	numpages = {6},
	year = {2022},
	month = {Sep},
	publisher = {American Physical Society},
	doi = {10.1103/PhysRevLett.129.137201},
	url = {https://link.aps.org/doi/10.1103/PhysRevLett.129.137201}
}

@article{Gonzalez2021,
	title = {{Efficient Electrical Spin Splitter Based on Nonrelativistic Collinear Antiferromagnetism}},
	author = {Gonz\'alez-Hern\'andez, Rafael and \ifmmode \check{S}\else \v{S}\fi{}mejkal, Libor and V\'yborn\'y, Karel and Yahagi, Yuta and Sinova, Jairo and Jungwirth, Tomas and \ifmmode \check{Z}\else \v{Z}\fi{}elezn\'y, Jakub},
	journal = {Phys. Rev. Lett.},
	volume = {126},
	issue = {12},
	pages = {127701},
	numpages = {6},
	year = {2021},
	month = {Mar},
	publisher = {American Physical Society},
	doi = {10.1103/PhysRevLett.126.127701},
	url = {https://link.aps.org/doi/10.1103/PhysRevLett.126.127701}
}

@article{Manchon2015,
	author = {Manchon, A. and Koo, H. C. and Nitta, J. and Frolov, S. M. and Duine, R. A.},
	date = {2015/09/01},
	doi = {10.1038/nmat4360},
	id = {Manchon2015},
	isbn = {1476-4660},
	journal = {Nature Materials},
	number = {9},
	pages = {871--882},
	title = {{New perspectives for {R}ashba spin--orbit coupling}},
	url = {https://doi.org/10.1038/nmat4360},
	volume = {14},
	year = {2015}}

@article{Miron2011,
	author = {Miron, Ioan Mihai and Garello, Kevin and Gaudin, Gilles and Zermatten, Pierre-Jean and Costache, Marius V. and Auffret, St{\'e}phane and Bandiera, S{\'e}bastien and Rodmacq, Bernard and Schuhl, Alain and Gambardella, Pietro},
	date = {2011/08/01},
	doi = {10.1038/nature10309},
	isbn = {1476-4687},
	journal = {Nature},
	number = {7359},
	pages = {189--193},
	title = {{Perpendicular switching of a single ferromagnetic layer induced by in-plane current injection}},
	url = {https://doi.org/10.1038/nature10309},
	volume = {476},
	year = {2011}}

@article{Miron2010,
	author = {M. Miron, Ioan and Gaudin, Gilles and Auffret, St{\'e}phane and Rodmacq, Bernard and Schuhl, Alain and Pizzini, Stefania and Vogel, Jan and Gambardella, Pietro},
	date = {2010/03/01},
	doi = {10.1038/nmat2613},
	isbn = {1476-4660},
	journal = {Nature Materials},
	number = {3},
	pages = {230--234},
	title = {{Current-driven spin torque induced by the Rashba effect in a ferromagnetic metal layer}},
	url = {https://doi.org/10.1038/nmat2613},
	volume = {9},
	year = {2010}}

@article{Salemi2019,
	author = {Salemi, Leandro and Berritta, Marco and Nandy, Ashis K. and Oppeneer, Peter M.},
	date = {2019/11/26},
	doi = {10.1038/s41467-019-13367-z},
	id = {Salemi2019},
	isbn = {2041-1723},
	journal = {Nature Communications},
	number = {1},
	pages = {5381},
	title = {{Orbitally dominated Rashba-Edelstein effect in noncentrosymmetric antiferromagnets}},
	url = {https://doi.org/10.1038/s41467-019-13367-z},
	volume = {10},
	year = {2019}}

@article{Zelezny2014,
	title = {{Relativistic N\'eel-Order Fields Induced by Electrical Current in Antiferromagnets}},
	author = {\ifmmode \check{Z}\else \v{Z}\fi{}elezn\'y, J. and Gao, H. and V\'yborn\'y, K. and Zemen, J. and Ma\ifmmode \check{s}\else \v{s}\fi{}ek, J. and Manchon, Aur\'elien and Wunderlich, J. and Sinova, Jairo and Jungwirth, T.},
	journal = {Phys. Rev. Lett.},
	volume = {113},
	issue = {15},
	pages = {157201},
	numpages = {5},
	year = {2014},
	month = {Oct},
	publisher = {American Physical Society},
	doi = {10.1103/PhysRevLett.113.157201},
	url = {https://link.aps.org/doi/10.1103/PhysRevLett.113.157201}
}

@article{Ciccarelli2016,
	author = {Ciccarelli, C. and Anderson, L. and Tshitoyan, V. and Ferguson, A. J. and Gerhard, F. and Gould, C. and Molenkamp, L. W. and Gayles, J. and {\v Z}elezn{\'y}, J. and {\v S}mejkal, L. and Yuan, Z. and Sinova, J. and Freimuth, F. and Jungwirth, T.},
	date = {2016/09/01},
	doi = {10.1038/nphys3772},
	id = {Ciccarelli2016},
	isbn = {1745-2481},
	journal = {Nature Physics},
	number = {9},
	pages = {855--860},
	title = {{Room-temperature spin--orbit torque in NiMnSb}},
	url = {https://doi.org/10.1038/nphys3772},
	volume = {12},
	year = {2016}}

@article{Kurebayashi2014,
	author = {Kurebayashi, H. and Sinova, Jairo and Fang, D. and Irvine, A. C. and Skinner, T. D. and Wunderlich, J. and Nov{\'a}k, V. and Campion, R. P. and Gallagher, B. L. and Vehstedt, E. K. and Z{\^a}rbo, L. P. and V{\'y}born{\'y}, K. and Ferguson, A. J. and Jungwirth, T.},
	date = {2014/03/01},
	doi = {10.1038/nnano.2014.15},
	id = {Kurebayashi2014},
	isbn = {1748-3395},
	journal = {Nature Nanotechnology},
	number = {3},
	pages = {211--217},
	title = {{An antidamping spin--orbit torque originating from the Berry curvature}},
	url = {https://doi.org/10.1038/nnano.2014.15},
	volume = {9},
	year = {2014}}

@article{Chernyshov2009,
	author = {Chernyshov, Alexandr and Overby, Mason and Liu, Xinyu and Furdyna, Jacek K. and Lyanda-Geller, Yuli and Rokhinson, Leonid P.},
	date = {2009/09/01},
	doi = {10.1038/nphys1362},
	id = {Chernyshov2009},
	isbn = {1745-2481},
	journal = {Nature Physics},
	number = {9},
	pages = {656--659},
	title = {{Evidence for reversible control of magnetization in a ferromagnetic material by means of spin--orbit magnetic field}},
	url = {https://doi.org/10.1038/nphys1362},
	volume = {5},
	year = {2009}}

@article{Fang2011,
	author = {Fang, D. and Kurebayashi, H. and Wunderlich, J. and V{\'y}born{\'y}, K. and Z{\^a}rbo, L. P. and Campion, R. P. and Casiraghi, A. and Gallagher, B. L. and Jungwirth, T. and Ferguson, A. J.},
	date = {2011/07/01},
	doi = {10.1038/nnano.2011.68},
	id = {Fang2011},
	isbn = {1748-3395},
	journal = {Nature Nanotechnology},
	number = {7},
	pages = {413--417},
	title = {{Spin--orbit-driven ferromagnetic resonance}},
	url = {https://doi.org/10.1038/nnano.2011.68},
	volume = {6},
	year = {2011}}

@article{Smejkal2022_1,
	title = {Beyond Conventional Ferromagnetism and Antiferromagnetism: A Phase with Nonrelativistic Spin and Crystal Rotation Symmetry},
	author = {\ifmmode \check{S}\else \v{S}\fi{}mejkal, Libor and Sinova, Jairo and Jungwirth, Tomas},
	journal = {Phys. Rev. X},
	volume = {12},
	issue = {3},
	pages = {031042},
	numpages = {16},
	year = {2022},
	month = {Sep},
	publisher = {American Physical Society},
	doi = {10.1103/PhysRevX.12.031042},
	url = {https://link.aps.org/doi/10.1103/PhysRevX.12.031042}
}

@article{Hayami2019,
	author = {Hayami ,Satoru and Yanagi ,Yuki and Kusunose ,Hiroaki},
	title = {Momentum-Dependent Spin Splitting by Collinear Antiferromagnetic Ordering},
	journal = {Journal of the Physical Society of Japan},
	volume = {88},
	number = {12},
	pages = {123702},
	year = {2019},
	doi = {10.7566/JPSJ.88.123702},
	
	URL = { 
	
	https://doi.org/10.7566/JPSJ.88.123702
	
	
	
	}
}

@article{Smejkal2022_3,
	title = {Giant and Tunneling Magnetoresistance in Unconventional Collinear Antiferromagnets with Nonrelativistic Spin-Momentum Coupling},
	author = {\ifmmode \check{S}\else \v{S}\fi{}mejkal, Libor and Hellenes, Anna Birk and Gonz\'alez-Hern\'andez, Rafael and Sinova, Jairo and Jungwirth, Tomas},
	journal = {Phys. Rev. X},
	volume = {12},
	issue = {1},
	pages = {011028},
	numpages = {11},
	year = {2022},
	month = {Feb},
	publisher = {American Physical Society},
	doi = {10.1103/PhysRevX.12.011028},
	url = {https://link.aps.org/doi/10.1103/PhysRevX.12.011028}
}

@Article{Reimers2024,
	author={Reimers, Sonka
	and Odenbreit, Lukas
	and {\v{S}}mejkal, Libor
	and Strocov, Vladimir N.
	and Constantinou, Procopios
	and Hellenes, Anna B.
	and Jaeschke Ubiergo, Rodrigo
	and Campos, Warlley H.
	and Bharadwaj, Venkata K.
	and Chakraborty, Atasi
	and others},
	title={Direct observation of altermagnetic band splitting in CrSb thin films},
	journal={Nature Communications},
	year={2024},
	month={Mar},
	day={08},
	volume={15},
	number={1},
	pages={2116},
	issn={2041-1723},
	doi={10.1038/s41467-024-46476-5},
	url={https://doi.org/10.1038/s41467-024-46476-5}
}

@article{Yarmohammadi2025,
	title = {Anisotropic light-tailored {RKKY} interaction in two-dimensional $d$-wave altermagnets},
	author = {Yarmohammadi, Mohsen and Z\"ulicke, Ulrich and Berakdar, Jamal and Linder, Jacob and Freericks, James K.},
	journal = {Phys. Rev. B},
	volume = {111},
	issue = {22},
	pages = {224412},
	numpages = {16},
	year = {2025},
	month = {Jun},
	publisher = {American Physical Society},
	doi = {10.1103/k3xb-8pts},
	url = {https://link.aps.org/doi/10.1103/k3xb-8pts}
	
}

@article{Lee2024,
	title={Magnetic impurities in an altermagnetic metal}, 
	author={Yu-Li Lee},
	year={2024},
	eprint={2312.15733},
	archivePrefix={arXiv},
	url={https://arxiv.org/abs/2312.15733}, 
}

@article{Yarmohammadi2026,
	title = {Spin polarization engineering in $d$-wave altermagnets},
	author = {Yarmohammadi, Mohsen and Berritta, Marco and Bukov, Marin and \ifmmode \check{S}\else \v{S}\fi{}mejkal, Libor and Linder, Jacob and Oppeneer, Peter M.},
	journal = {Phys. Rev. B},
	volume = {113},
	issue = {6},
	pages = {L060403},
	numpages = {6},
	year = {2026},
	month = {Feb},
	publisher = {American Physical Society},
	doi = {10.1103/xt23-9pnv},
	url = {https://link.aps.org/doi/10.1103/xt23-9pnv}
}

@article{SM,
	journal={See Supplemental Material at [URL will be inserted by publisher] for providing ... }
}

@article{Smejkal2022_2,
  title = {Emerging Research Landscape of Altermagnetism},
  author = {\ifmmode \check{S}\else \v{S}\fi{}mejkal, Libor and Sinova, Jairo and Jungwirth, Tomas},
  journal = {Phys. Rev. X},
  volume = {12},
  issue = {4},
  pages = {040501},
  numpages = {27},
  year = {2022},
  month = {Dec},
  publisher = {American Physical Society},
  doi = {10.1103/PhysRevX.12.040501},
  url = {https://link.aps.org/doi/10.1103/PhysRevX.12.040501}
}

@article{Yuan2020,
  title = {Giant momentum-dependent spin splitting in centrosymmetric low-$Z$ antiferromagnets},
  author = {Yuan, Lin-Ding and Wang, Zhi and Luo, Jun-Wei and Rashba, Emmanuel I. and Zunger, Alex},
  journal = {Phys. Rev. B},
  volume = {102},
  issue = {1},
  pages = {014422},
  numpages = {13},
  year = {2020},
  month = {Jul},
  publisher = {American Physical Society},
  doi = {10.1103/PhysRevB.102.014422},
  url = {https://link.aps.org/doi/10.1103/PhysRevB.102.014422}
}

@article{Neel1953,
  title = {Some New Results on Antiferromagnetism and Ferromagnetism},
  author = {N\'eel, Louis},
  journal = {Rev. Mod. Phys.},
  volume = {25},
  issue = {1},
  pages = {58--63},
  numpages = {0},
  year = {1953},
  month = {Jan},
  publisher = {American Physical Society},
  doi = {10.1103/RevModPhys.25.58},
  url = {https://link.aps.org/doi/10.1103/RevModPhys.25.58}
}

@article{Rashba1960,
author="RASHBA, E.",
title="Properties of semiconductors with an extremum loop. I. Cyclotron and combinational Resonance in a magnetic field perpendicular to the plane of the loop",
journal="Sov. Phys.-Solid State",
year="1960",
volume="2",
pages="1109",
URL="https://cir.nii.ac.jp/crid/1571698600346713472"
}

@article{Dresselhaus1955,
  title = {Spin-Orbit Coupling Effects in Zinc Blende Structures},
  author = {Dresselhaus, G.},
  journal = {Phys. Rev.},
  volume = {100},
  issue = {2},
  pages = {580--586},
  numpages = {0},
  year = {1955},
  month = {Oct},
  publisher = {American Physical Society},
  doi = {10.1103/PhysRev.100.580},
  url = {https://link.aps.org/doi/10.1103/PhysRev.100.580}
}

@article{Yuan2021,
  title = {Prediction of low-Z collinear and noncollinear antiferromagnetic compounds having momentum-dependent spin splitting even without spin-orbit coupling},
  author = {Yuan, Lin-Ding and Wang, Zhi and Luo, Jun-Wei and Zunger, Alex},
  journal = {Phys. Rev. Mater.},
  volume = {5},
  issue = {1},
  pages = {014409},
  numpages = {24},
  year = {2021},
  month = {Jan},
  publisher = {American Physical Society},
  doi = {10.1103/PhysRevMaterials.5.014409},
  url = {https://link.aps.org/doi/10.1103/PhysRevMaterials.5.014409}
}

@article{Yuan2024,
  title = {Nonrelativistic Spin Splitting at the Brillouin Zone Center in Compensated Magnets},
  author = {Yuan, Lin-Ding and Georgescu, Alexandru B. and Rondinelli, James M.},
  journal = {Phys. Rev. Lett.},
  volume = {133},
  issue = {21},
  pages = {216701},
  numpages = {8},
  year = {2024},
  month = {Nov},
  publisher = {American Physical Society},
  doi = {10.1103/PhysRevLett.133.216701},
  url = {https://link.aps.org/doi/10.1103/PhysRevLett.133.216701}
}

@article{Zhang2025,
  title = {Strain-induced nonrelativistic altermagnetic spin splitting effect},
  author = {Zhang, Wancheng and Zheng, Mingkun and Liu, Yong and Zhang, Zhenhua and Xiong, Rui and Lu, Zhihong},
  journal = {Phys. Rev. B},
  volume = {112},
  issue = {2},
  pages = {024415},
  numpages = {11},
  year = {2025},
  month = {Jul},
  publisher = {American Physical Society},
  doi = {10.1103/8zlt-mlms},
  url = {https://link.aps.org/doi/10.1103/8zlt-mlms}
}

@article{Bandyopadhyay2025,
  title = {Designing nonrelativistic spin splitting in oxide perovskites},
  author = {Bandyopadhyay, Subhadeep and Picozzi, Silvia and Bhowal, Sayantika},
  journal = {Phys. Rev. B},
  volume = {112},
  issue = {6},
  pages = {064405},
  numpages = {15},
  year = {2025},
  month = {Aug},
  publisher = {American Physical Society},
  doi = {10.1103/1r6k-s46h},
  url = {https://link.aps.org/doi/10.1103/1r6k-s46h}
}

@article{Yarmohammadi2025_2,
	title = {Cavity-assisted magnetization switching in a quantum spin-phonon chain},
	author = {Yarmohammadi, Mohsen and Oppeneer, Peter M. and Freericks, James K.},
	journal = {Phys. Rev. B},
	volume = {112},
	issue = {9},
	pages = {094445},
	numpages = {10},
	year = {2025},
	month = {Sep},
	publisher = {American Physical Society},
	doi = {10.1103/88tw-h78r},
	url = {https://link.aps.org/doi/10.1103/88tw-h78r}
}

@ARTICLE{Rajpurohit2024,
       author = {{Rajpurohit}, Sangeeta and {Karaalp}, Revsen and {Ping}, Yuan and {Tan}, Liang Z. and {Ogitsu}, Tadashi and {Bl{\"o}chl}, Peter E.},
        title = "{Optical control of spin-splitting in an altermagnet}",
         year = 2024,
        month = sep,
          eid = {arXiv:2409.17718},
        pages = {arXiv:2409.17718},
          doi = {10.48550/arXiv.2409.17718},
       adsurl = {https://ui.adsabs.harvard.edu/abs/2024arXiv240917718R}
}

@ARTICLE{Guo2023,
       author = {{Guo}, Yaqian and {Liu}, Hui and {Janson}, Oleg and {Fulga}, Ion Cosma and {van den Brink}, Jeroen and {Facio}, Jorge I.},
        title = "{Spin-split collinear antiferromagnets: A large-scale ab-initio study}",
      journal = {Materials Today Physics},
         year = 2023,
        month = mar,
       volume = {32},
          eid = {100991},
        pages = {100991},
          doi = {10.1016/j.mtphys.2023.100991},
archivePrefix = {arXiv},
       adsurl = {https://ui.adsabs.harvard.edu/abs/2023MTPhy..3200991G}
}

@article{Xian2019,
author = {Xian Li  and Tian Qiu  and Jiahao Zhang  and Edoardo Baldini  and Jian Lu  and Andrew M. Rappe  and Keith A. Nelson },
title = {Terahertz field--induced ferroelectricity in quantum paraelectric {S}r{T}i{O}$_3$},
journal = {Science},
volume = {364},
number = {6445},
pages = {1079-1082},
year = {2019},
doi = {10.1126/science.aaw4913},
URL = {https://www.science.org/doi/abs/10.1126/science.aaw4913}}

@article{Cavalleri2018,
	author = {Andrea Cavalleri},
	doi = {10.1080/00107514.2017.1406623},
	journal = {Contemporary Physics},
	number = {1},
	pages = {31--46},
	publisher = {Taylor \& Francis},
	title = {Photo-induced superconductivity},
	url = {https://doi.org/10.1080/00107514.2017.1406623},
	volume = {59},
	year = {2018}
}

@ARTICLE{Haque2025,
       author = {{Haque}, Sheikh Rubaiat Ul and {Cross}, Martin J. and {Rajpurohit}, Sangeeta and {Haber}, Jonah B. and {Ciccarino}, Christopher J. and {Zimmerman}, Alexandra C. and {Sealey}, Isabelle J. and {Kulichenko}, Vadym and {Tie}, Monique and {Wang}, Huaiyu and {Philip}, Sharon S. and {Seo}, Choongwon and {Koralek}, Jake D. and {Balicas}, Luis and {Ozerov}, Mykhaylo and {Smirnov}, Dmitry and {Tan}, Liang Z. and {da Jornada}, Felipe H. and {Ogitsu}, Tadashi and {Hoffmann}, Matthias C. and {Heinz}, Tony F. and {Lindenberg}, Aaron M.},
        title = "{Terahertz field-induced giant symmetry modulations in a van der Waals antiferromagnet}",
         year = 2025,
        month = oct,
          eid = {arXiv:2510.00605},
        pages = {arXiv:2510.00605},
          doi = {10.48550/arXiv.2510.00605},
       adsurl = {https://ui.adsabs.harvard.edu/abs/2025arXiv251000605H}
}

@article{Disa2023,
	author = {Disa, A. S. and Curtis, J. and Fechner, M. and Liu, A. and von Hoegen, A. and F{\"o}rst, M. and Nova, T. F. and Narang, P. and Maljuk, A. and Boris, A. V. and Keimer, B. and Cavalleri, A.},
	date = {2023/05/01},
	doi = {10.1038/s41586-023-05853-8},
	id = {Disa2023},
	isbn = {1476-4687},
	journal = {Nature},
	number = {7959},
	pages = {73--78},
	title = {Photo-induced high-temperature ferromagnetism in YTiO3},
	url = {https://doi.org/10.1038/s41586-023-05853-8},
	volume = {617},
	year = {2023}
}

@article{Liu2022,
  title = {Spin-Group Symmetry in Magnetic Materials with Negligible Spin-Orbit Coupling},
  author = {Liu, Pengfei and Li, Jiayu and Han, Jingzhi and Wan, Xiangang and Liu, Qihang},
  journal = {Phys. Rev. X},
  volume = {12},
  issue = {2},
  pages = {021016},
  numpages = {19},
  year = {2022},
  month = {Apr},
  publisher = {American Physical Society},
  doi = {10.1103/PhysRevX.12.021016},
  url = {https://link.aps.org/doi/10.1103/PhysRevX.12.021016}
}

@ARTICLE{Giannozzi2009,
       author = {{Giannozzi}, Paolo and {Baroni}, Stefano and {Bonini}, Nicola and {Calandra}, Matteo and {Car}, Roberto and {Cavazzoni}, Carlo and {Ceresoli}, Davide and {Chiarotti}, Guido L. and {Cococcioni}, Matteo and {Dabo}, Ismaila and {Dal Corso}, Andrea and {de Gironcoli}, Stefano and {Fabris}, Stefano and {Fratesi}, Guido and {Gebauer}, Ralph and {Gerstmann}, Uwe and {Gougoussis}, Christos and {Kokalj}, Anton and {Lazzeri}, Michele and {Martin-Samos}, Layla and {Marzari}, Nicola and {Mauri}, Francesco and {Mazzarello}, Riccardo and {Paolini}, Stefano and {Pasquarello}, Alfredo and {Paulatto}, Lorenzo and {Sbraccia}, Carlo and {Scandolo}, Sandro and {Sclauzero}, Gabriele and {Seitsonen}, Ari P. and {Smogunov}, Alexander and {Umari}, Paolo and {Wentzcovitch}, Renata M.},
        title = "{QUANTUM ESPRESSO: a modular and open-source software project for quantum simulations of materials}",
      journal = {Journal of Physics Condensed Matter},
         year = 2009,
        month = sep,
       volume = {21},
       number = {39},
          eid = {395502},
        pages = {395502},
          doi = {10.1088/0953-8984/21/39/395502},
       adsurl = {https://ui.adsabs.harvard.edu/abs/2009JPCM...21M5502G}
}

@ARTICLE{Giannozzi2017,
       author = {{Giannozzi}, P. and {Andreussi}, O. and {Brumme}, T. and {Bunau}, O. and {Buongiorno Nardelli}, M. and {Calandra}, M. and {Car}, R. and {Cavazzoni}, C. and {Ceresoli}, D. and {Cococcioni}, M. and {Colonna}, N. and {Carnimeo}, I. and {Dal Corso}, A. and {de Gironcoli}, S. and {Delugas}, P. and {DiStasio}, Jr., R.~A. and {Ferretti}, A. and {Floris}, A. and {Fratesi}, G. and {Fugallo}, G. and {Gebauer}, R. and {Gerstmann}, U. and {Giustino}, F. and {Gorni}, T. and {Jia}, J. and {Kawamura}, M. and {Ko}, H.-Y. and {Kokalj}, A. and {K{\"u}{\c{c}}{\"u}kbenli}, E. and {Lazzeri}, M. and {Marsili}, M. and {Marzari}, N. and {Mauri}, F. and {Nguyen}, N.~L. and {Nguyen}, H.-V. and {Otero-de-la-Roza}, A. and {Paulatto}, L. and {Ponc{\'e}}, S. and {Rocca}, D. and {Sabatini}, R. and {Santra}, B. and {Schlipf}, M. and {Seitsonen}, A.~P. and {Smogunov}, A. and {Timrov}, I. and {Thonhauser}, T. and {Umari}, P. and {Vast}, N. and {Wu}, X. and {Baroni}, S.},
        title = "{Advanced capabilities for materials modelling with Quantum ESPRESSO}",
      journal = {Journal of Physics Condensed Matter},
         year = 2017,
        month = nov,
       volume = {29},
       number = {46},
          eid = {465901},
        pages = {465901},
          doi = {10.1088/1361-648X/aa8f79},
       adsurl = {https://ui.adsabs.harvard.edu/abs/2017JPCM...29T5901G}
}

@article{Perdew1996,
    author = {Perdew, John P. and Ernzerhof, Matthias and Burke, Kieron},
    title = {Rationale for mixing exact exchange with density functional approximations},
    journal = {The Journal of Chemical Physics},
    volume = {105},
    number = {22},
    pages = {9982-9985},
    year = {1996},
    month = {12},
    issn = {0021-9606},
    doi = {10.1063/1.472933},
    url = {https://doi.org/10.1063/1.472933},
}

@article{Vaclavkova2020,
doi = {10.1088/2053-1583/ab93e3},
url = {https://doi.org/10.1088/2053-1583/ab93e3},
year = {2020},
month = {jun},
publisher = {IOP Publishing},
volume = {7},
number = {3},
pages = {035030},
author = {Vaclavkova, Diana and Delhomme, Alex and Faugeras, Clément and Potemski, Marek and Bogucki, Aleksander and Suffczyński, Jan and Kossacki, Piotr and Wildes, Andrew R and Grémaud, Benoit and Saúl, Andrés},
title = {Magnetoelastic interaction in the two-dimensional magnetic material MnPS3 studied by first principles calculations and Raman experiments},
journal = {2D Materials}
}

@article{Rao2024,
author = {Rao, Rahul and Selhorst, Ryan and Siebenaller, Ryan and Giordano, Andrea N. and Conner, Benjamin S. and Rowe, Emmanuel and Susner, Michael A.},
title = {Mode-Selective Spin–Phonon Coupling in van der Waals Antiferromagnets},
journal = {Advanced Physics Research},
volume = {3},
number = {6},
pages = {2300153},
doi = {https://doi.org/10.1002/apxr.202300153},
url = {https://advanced.onlinelibrary.wiley.com/doi/abs/10.1002/apxr.202300153},
year = {2024}
}

@article{Sie2019,
	author = {Sie, Edbert J. and Nyby, Clara M. and Pemmaraju, C. D. and Park, Su Ji and Shen, Xiaozhe and Yang, Jie and Hoffmann, Matthias C. and Ofori-Okai, B. K. and Li, Renkai and Reid, Alexander H. and Weathersby, Stephen and Mannebach, Ehren and Finney, Nathan and Rhodes, Daniel and Chenet, Daniel and Antony, Abhinandan and Balicas, Luis and Hone, James and Devereaux, Thomas P. and Heinz, Tony F. and Wang, Xijie and Lindenberg, Aaron M.},
	date = {2019/01/01},
	doi = {10.1038/s41586-018-0809-4},
	id = {Sie2019},
	isbn = {1476-4687},
	journal = {Nature},
	number = {7737},
	pages = {61--66},
	title = {An ultrafast symmetry switch in a Weyl semimetal},
	url = {https://doi.org/10.1038/s41586-018-0809-4},
	volume = {565},
	year = {2019}
}

@ARTICLE{Wrzos2025,
       author = {{Wrzos}, Kamil and {Birowska}, Magdalena and {Rybak}, Milosz},
        title = "{Symmetry-Breaking Phenomena in MnPS3/TMDC Heterostructures: Non-relativistic Spin Splitting, Altermagnetism and Spin-Valley Effects}",
         year = 2025,
        month = nov,
          eid = {arXiv:2511.22209},
        pages = {arXiv:2511.22209},
          doi = {10.48550/arXiv.2511.22209},
       adsurl = {https://ui.adsabs.harvard.edu/abs/2025arXiv251122209W}
}

@ARTICLE{Wang2026,
       author = {{Wang}, Chenyu and {Wang}, Yaxian and {Meng}, Sheng},
        title = "{Ultrafast altermagnetophononics}",
      journal = {arXiv e-prints},
         year = 2026,
        month = jul,
          eid = {arXiv:2607.13863},
        pages = {arXiv:2607.13863},
          doi = {10.48550/arXiv.2607.13863},
archivePrefix = {arXiv},
       eprint = {2607.13863},
 primaryClass = {cond-mat.mtrl-sci},
       adsurl = {https://ui.adsabs.harvard.edu/abs/2026arXiv260713863W}
}

@article{Fiebig2005,
	doi = {10.1364/JOSAB.22.000096},
	journal = {J. Opt. Soc. Am. B},
	month = {Jan},
	number = {1},
	pages = {96--118},
	publisher = {Optica Publishing Group},
	title = {Second-harmonic generation as a tool for studying electronic and magnetic structures of crystals: review},
	url = {https://opg.optica.org/josab/abstract.cfm?URI=josab-22-1-96},
	volume = {22},
	year = {2005}
}

@article{Bloechl1994,
  title = {Projector augmented-wave method},
  author = {Bl\"ochl, P. E.},
  journal = {Phys. Rev. B},
  volume = {50},
  issue = {24},
  pages = {17953--17979},
  numpages = {0},
  year = {1994},
  month = {Dec},
  publisher = {American Physical Society},
  doi = {10.1103/PhysRevB.50.17953},
  url = {https://link.aps.org/doi/10.1103/PhysRevB.50.17953}
}

\let\addcontentsline\oldaddcontentsline

	\onecolumngrid
	\clearpage

	{\allowdisplaybreaks
		
		\begin{center}
			\textbf{\large \vskip0mm Supplemental Materials for ``Symmetry-selective nonrelativistic spin splitting in antiferromagnets driven by coherent phonons''}
            \vskip3.5mm
			Sangeeta Rajpurohit,$^{1}$ Mohsen Yarmohammadi$^{2}$, Sheikh Rubaiat Ul\\ Haque$^{3,4,5}$, Tony F. Heinz$^{3,5,6}$, Aaron M. Lindenberg,$^{4,5,6}$ and Tadashi Ogitsu$^{1}$\vskip1mm
			\small $^1$\textit{Material Science Division, Lawrence Livermore National Laboratory, CA 94550, USA}\\
            \small $^2$\textit{Department of Physics, Georgetown University, Washington DC 20057, USA}\\
            \small $^3$\textit{Department of Applied Physics, Stanford University, Stanford, CA 94305, USA}\\
            \small $^4$\textit{Department of Materials Science and Engineering, Stanford University, Stanford, CA 94305, USA}\\
            \small $^5$\textit{Stanford Institute for Materials and Energy Sciences, \\SLAC National Accelerator Laboratory, Menlo Park, CA 94025, USA}\\
            \small $^6$\textit{Stanford PULSE Institute, SLAC National Accelerator Laboratory, Menlo Park, CA 94025, USA}\\
            (Dated: \today)
		\end{center}

		\setcounter{equation}{0}
		\setcounter{figure}{0}
		\setcounter{section}{0}
		\setcounter{page}{1}
		\makeatletter

		\setcounter{equation}{0}
		\renewcommand{\theequation}{S\arabic{equation}}
		\setcounter{figure}{0}
		\renewcommand{\thefigure}{S\arabic{figure}}
		\setcounter{section}{0}
		\renewcommand{\thesection}{S\arabic{section}}
		\setcounter{table}{0}
		\renewcommand{\thetable}{S\arabic{table}}

        \vspace{0.5cm}
        
\addtocontents{toc}{\protect\setcounter{tocdepth}{2}}
\tableofcontents

\vspace{0.5cm}

\section{S1 Effect of Hubbard $U$}
\label{si:Hubbard_u}

The phonon-induced spin-splitting presented in the main text is robust against the choice of Hubbard $U$. We repeated
the frozen-phonon calculations for the $A_u$ and $B_u$ modes at $A=0.05$~\AA{} using $U=1.25$, $2.5$, $4.0$, and $5.0$~eV.
As shown in Fig.~\ref{fig:fig_s1}(a), the spin splitting changes only moderately with $U$ and remains of the same order of magnitude
throughout this range. For example, the $B_u$ conduction splitting decreases from $64$ to $58$~meV, while the $A_u$
valence splitting decreases from $39$ to $29$~meV. Importantly, the symmetry of the splitting remains unchanged:
the $A_u$ mode retains a finite $s$-wave splitting at $\Gamma$, whereas the $B_u$ mode retains the $d$-wave node at
$\Gamma$ for all values of $U$ [Figs.~\ref{fig:fig_s2}--\ref{fig:fig_s3}]. These results confirm that the
phonon-induced spin splitting and its symmetry are not sensitive to the chosen value of $U$.

\begin{figure}[b]
\centering
\includegraphics[width=0.6\linewidth]{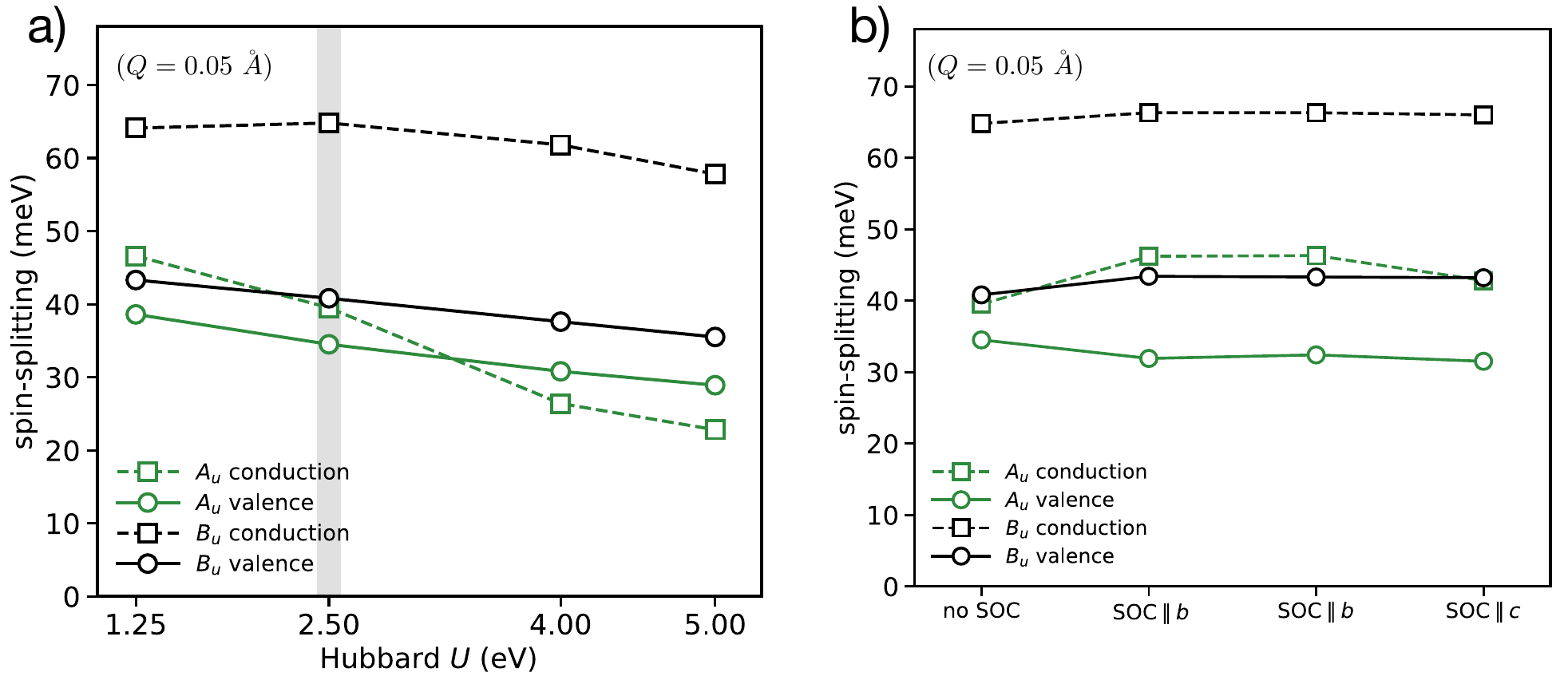}
\caption{Maximum valence (circles) and conduction (squares)
spin-splitting of the $A_u$ (green) and $B_u$ (black) modes at
$Q=0.05$~\AA, versus (a)~Hubbard $U$ ($U = 2.5$~eV shaded) and
(b)~SOC (none, and N\'eel $\parallel a/b/c$). The effect is nearly
independent of $U$, SOC, and N\'eel orientation.}
\label{fig:fig_s1}
\end{figure}

\section{S2. Effect of spin--orbit coupling}
\label{si:soc}

Here we show the phonon-induced NRSS is fundamentally non-relativistic:
its $s$- and $d$-wave symmetry is set by the spin space group and survives the
inclusion of SOC, which enters only as a weak perturbation ($\sim 10\%$ for the
$s$-wave splitting; a few-meV gap at the otherwise symmetry-protected $d$-wave
node). To verify that the phonon-induced spin splitting and its $s/d$-wave symmetry are robust
against relativistic effects, we repeated the frozen-phonon calculations for the $A_u$
and $B_u$ modes ($A=0.05$~\AA) including spin--orbit coupling (SOC). We performed
noncollinear DFT$+U$ calculations using fully relativistic PAW pseudopotentials, with
the same PBE functional, Hubbard $U{=}2.5$~eV, and plane-wave cutoffs of $80$~Ry for the
wavefunctions and $320$~Ry for the charge density. A denser $6\times6\times4$ Monkhorst--Pack grid
was used to resolve the small relativistic effects. Since spin is no longer a good quantum
number in the presence of SOC, we characterize each band by its spin expectation value
$\langle\mathbf{S}\rangle$ and determine the spin splitting from its projection along
the N\'eel axis $\hat{\mathbf n}$. To investigate the dependence on the moment orientation, 
we constrained the N\'eel vector along three orthogonal directions: the two in-plane axes
($a$, $b$) and the out-of-plane direction ($c$).

\begin{figure}[t]
\centering
\includegraphics[width=0.93\linewidth]{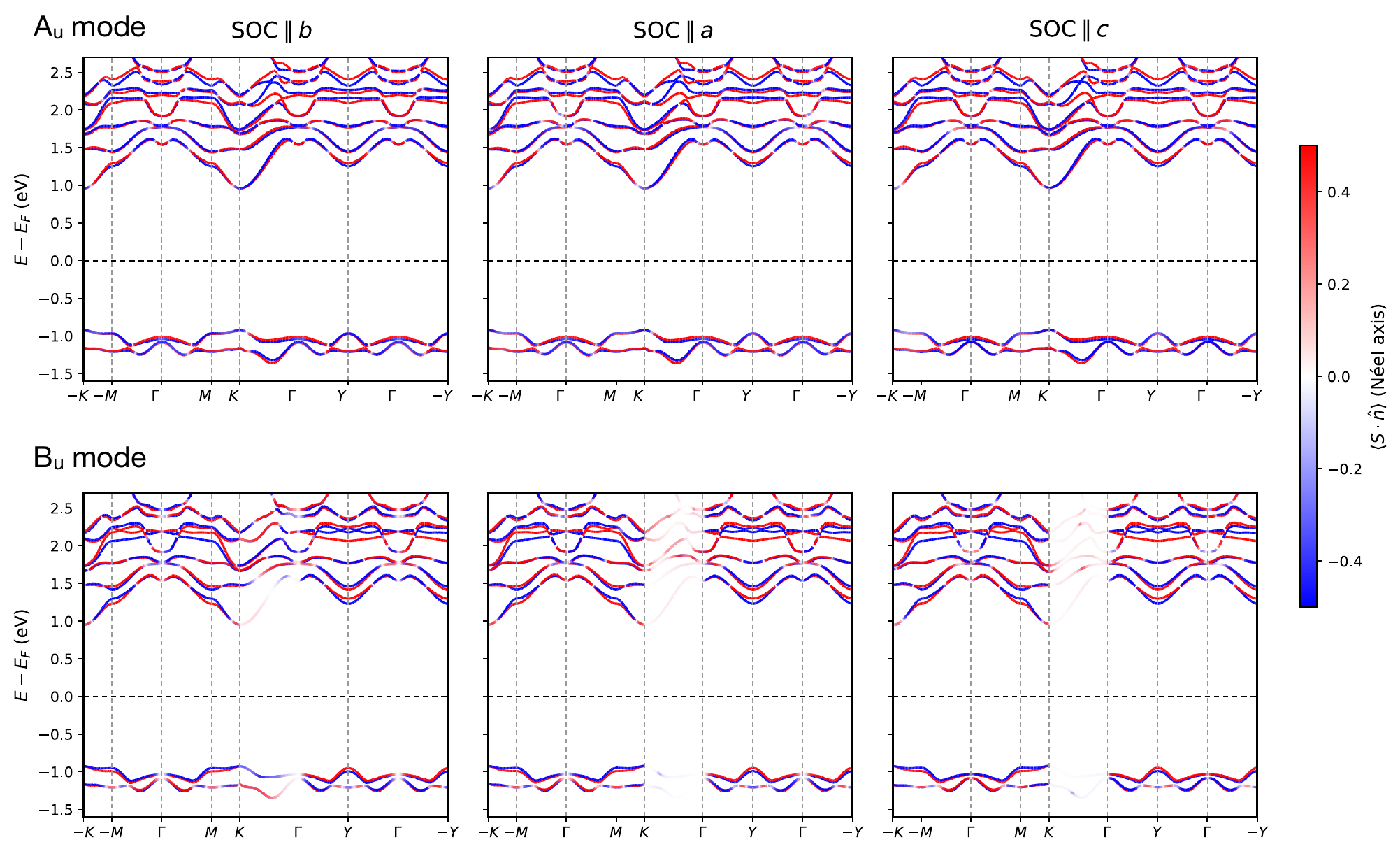}
\caption{Fully relativistic (SOC) band structures of MnPS$_3$ under the two
IR-mode distortions ($Q=0.05$~\AA), coloured by the spin projection on the
N\'eel axis $\langle \mathbf S\cdot\hat{\mathbf n}\rangle$ (red $+$, blue $-$),
for the N\'eel vector along $b$, $a$, $c$ (columns). Top row: the $s$-wave ($A_u$) mode
stays split through $\Gamma$. Bottom row: the $d$-wave ($B_u$) modes show the sign-changing
texture with a near-node at $\Gamma$.}
\label{fig:fig_s2}
\end{figure}
In the undistorted structure, every band remains Kramers-degenerate at all $\mathbf{k}$ for all
three N\'eel orientations (splitting $\le 1.5$~meV), consistent with the $P\mathcal{T}$ symmetry
that is retained with SOC. This confirms that the equilibrium spin degeneracy is not lifted by
SOC, and provides the zero reference for the distorted cases. Figures~\ref{fig:fig_s1} and~\ref{fig:fig_s3} (top set) compare the
zone-center splitting $\Delta(\Gamma)$ of the top and bottom of the valence and conduction band without and with
SOC. The $s$-wave $A_u$ mode retains the large spins-splitting at Gamma $\Delta(\Gamma)$, with $34.5$~meV without
SOC vs.\ $31$-$33$~meV with SOC, leaving the $s$-wave character intact.

The $d$-wave spin splitting in the $B_u$-distorted structure has $\Delta(\Gamma)=0$ exactly without SOC. 
The node is protected by the sublattice-connecting spin-space operation $[U\|\sigma_b]$ (equivalently $\sigma_b\mathcal{T}$),
which is exact only when spin and lattice are decoupled. With SOC, the node opens by a few meV
when the moments lie in-plane  ($\sim 9$~meV for $B_u$), while it remains almost negligible ($0.5$~meV ) when
the moments point along $c$, where a relativistic remnant of $\sigma_b\mathcal{T}$ survives. Crucially,
this SOC-induced node splitting is much smaller than the $d$-wave antinode, so the sign-changing $d$-wave texture is preserved; SOC merely opens a small relativistic
gap at $\Gamma$ (Fig.~\ref{fig:fig_3}, bottom set).

\begin{figure}[h]
\centering
\includegraphics[width=0.9\linewidth]{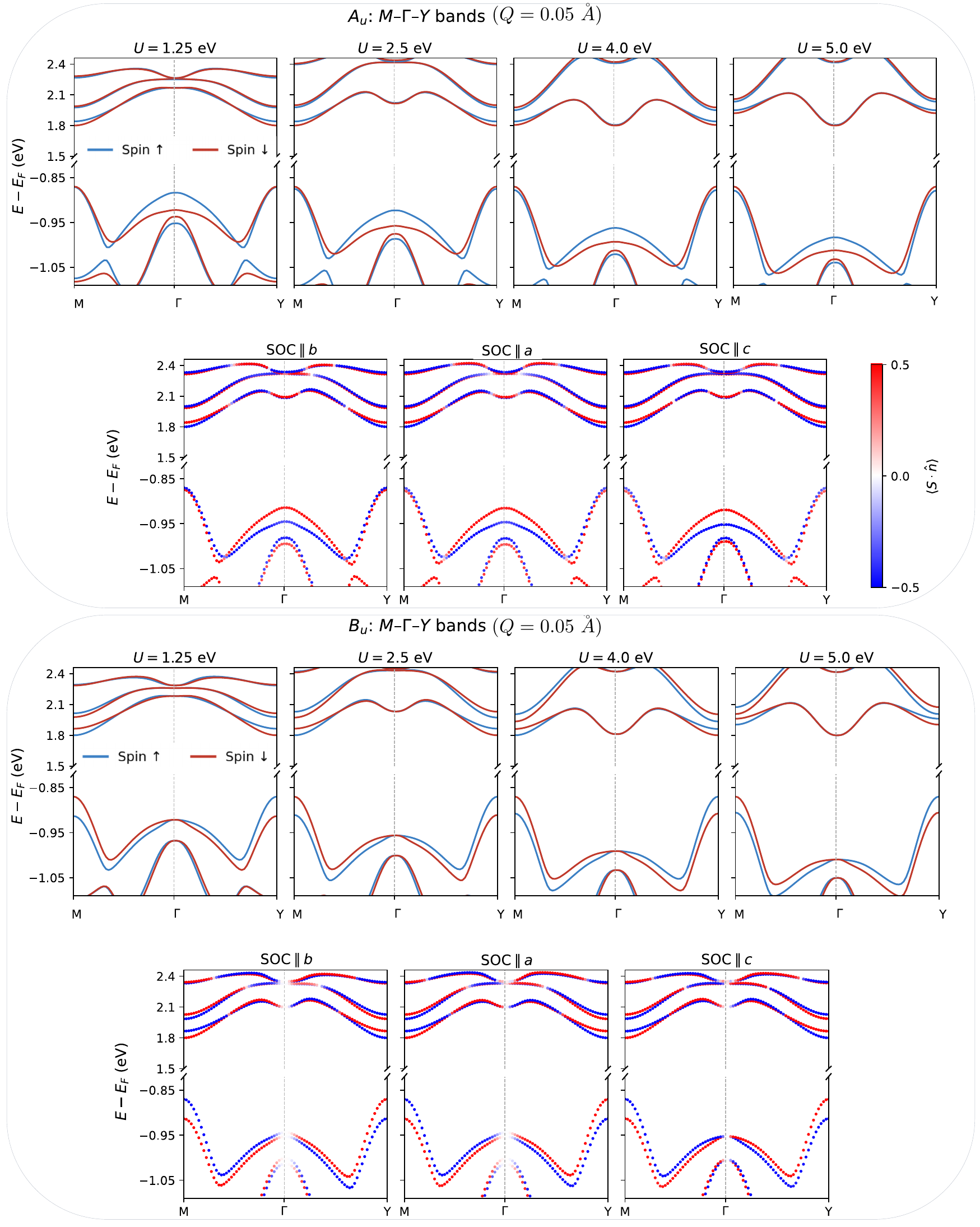}
\caption{(Top set) Spin-resolved band structure of the $A_u$-distorted state of
    MnPS$_3$ along $M$--$\Gamma$--$Y$ at $A = 0.05$~\AA. Top row: dependence
    on the Hubbard $U$ ($U = 1.25$, $2.5$, $4.0$, $5.0$~eV, without SOC),
    with spin-up (blue) and spin-down (red) bands. Bottom row: fully
    relativistic bands with the N\'eel vector along $b$, $a$, and $c$,
    colored by the spin projection $\langle S\cdot\hat n\rangle$. The
    topmost valence band is spin-split at $\Gamma$ in every case, and the
    $s$-wave splitting is robust to $U$, SOC, and N\'eel orientation. (Bottom set) Spin-resolved band structure of the $B_u$-distorted state of
    MnPS$_3$ along $M$--$\Gamma$--$Y$ at $A = 0.05$~\AA. Top row: dependence
    on the Hubbard $U$ ($U = 1.25$, $2.5$, $4.0$, $5.0$~eV, without SOC),
    with spin-up (blue) and spin-down (red) bands. Bottom row: fully
    relativistic bands with the N\'eel vector along $b$, $a$, and $c$,
    colored by the spin projection $\langle S\cdot\hat n\rangle$. The
    topmost valence band is spin-degenerate at $\Gamma$ (the $d$-wave node)
    and splits away from $\Gamma$ in every case; the node is preserved
    across all $U$ and only slightly lifted by SOC.}
\label{fig:fig_s3}
\end{figure}

    }

\end{document}